# A model based DC analysis of SiPM breakdown voltages

Ferenc Nagy, Gyula Hegyesi, Gábor Kalinka, József Molnár

## Abstract

A new method to determine the breakdown voltage of SiPMs is presented. It is based on a DC model which describes the breakdown phenomenon by distinct avalanche turn-on ($V_{01}$) and turn off ($V_{10}$) voltages. It is shown that traditional DC methods relying on the analysis of reverse current-voltage curves measure a value either near $V_{01}$ or between $V_{01}$ and $V_{10}$ while $V_{10}$ results by complex gain-voltage measurements. The proposed method reveals how the microcell population distributes around $V_{01}$. It is found that if this distribution is assumed to be normal, then both $V_{01}$ and $V_{10}$ of the SiPM can readily be extracted from current-voltage curves. Measurements are in good agreement with the theoretical model.

## 1. Introduction

A particle detector system design often demands that the various sensors of the system have similar parameters.

Due to uncertainties in the sensor production process, the parameters of the produced sensors always exhibit a distribution. This is also true for silicon photomultiplier (SiPM) production, where the least controllable parameter is the breakdown voltage. In fact the breakdown voltage values spread out in a wide range from SiPM to SiPM, especially if they are from different wafers or lots. [1]

There are two widely used approaches to characterize and select the sensors: taking a single photon spectrum (SPS) or a current-voltage curve (I-V) of the SiPMs. The first approach is sophisticated and time consuming while the second one, a simple DC measurement, is much faster.

A further limitation of the SPS approach is its sensitivity to the dark count rate. For example, for irradiated SiPMs (with gamma, neutron, proton etc.), the dark count rate may significantly increase. The smearing of the photoelectron peaks in the single photon spectrum makes them indistinguishable [2], [3], thus one can rely only on the I-V curve to characterize these devices.

In this article we focus on the I-V curve based (DC) characterization of SiPMs.

Traditionally, the breakdown voltage of an SiPM is determined from the logarithm of its current-voltage curve. There are several known methods such as the "tangent", "relative derivative" [4], [5], "inverse relative derivative" [3] and "second derivative" [6] methods. Besides these logarithm-based techniques, the "parabolic fitting" [7], [8] method processes the original current-voltage curve .

Table 1 summarizes these DC methods with instructive figures using a real I-V dataset.

The methods shown in Table 1 are either heuristic or based on the simple assumption that all microcells in an SiPM have the same breakdown voltage value. Moreover, for a given SiPM device the breakdown voltage values measured by the single photon spectrum and DC methods will differ [9]. To give an explanation for this phenomenon, a DC model is presented in Section 2. Based on the new model, we then propose what we call the 3$^{rd}$ derivative method, which can determine two distinct breakdown voltages of an SiPM from an I-V curve. The one with the lower value is the breakdown voltage that a complex single photon spectrum measurement would yield.

Theoretically in DC measurements both characteristic voltages appear, but since the known DC methods rely on the region of the I-V curve far from its corner point, their measured breakdown voltages are more related to the turn-on $V_{01}$ voltage. The 3$^{rd}$ derivative method concentrates on the region around the corner of the I-V curve, thus it reveals both $V_{01}$ and $V_{10}$ voltages of the SiPM.

The DC model and the 3$^{rd}$ derivative method are justified by measurements in Section 4 where we also test the reliability of our new method and compare it with all described DC methods.

**Table 1:** Known DC methods to determine the SiPM breakdown voltage from a real I-V curve

| | | | |
|---|---|---|---|
| **Tangent** | Linear fitted "baseline" and tangent drawn to $ln(I)$ | Intercept of tangent and the "baseline" | 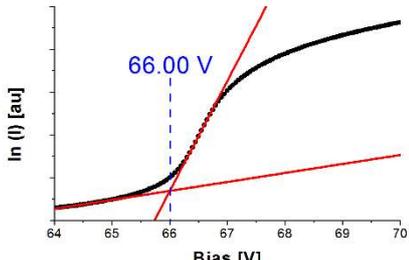 |
| **Relative derivative** | $\frac{d}{dV} ln(I) = I'/I$ | Position of the maximum | 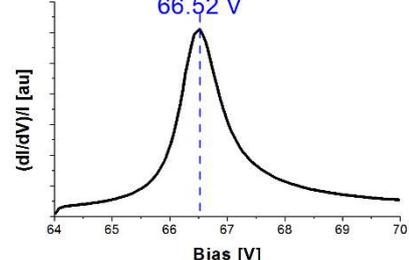 |
| **"Inverse" relative derivative** | $1 / \frac{d}{dV} ln(I) = I/I'$ | Intercept of the x-axis and the fitted line | 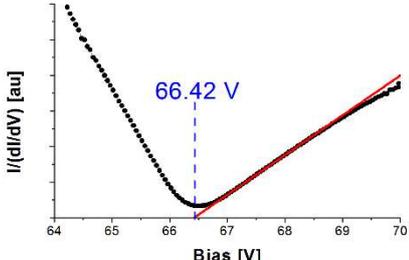 |
| **Second derivative** | $\frac{d^2}{dV^2} ln(I)$ | Position of the maximum | 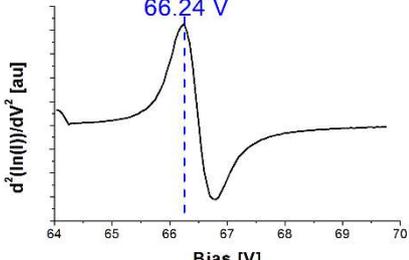 |
| **Parabolic fitting** | Linear fitted "baseline" and parabola fitted to I | Intercept of the fitted parabola and the "baseline" on semi-log scale | 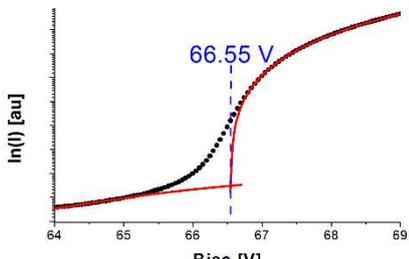 |

## 2. DC model and the 3$^{rd}$ derivative method

Just above the SiPM breakdown the presence of crosstalk and afterpulse is negligible. Thus in this voltage region a quadratic approximation can be used for the current-voltage function of a single cell: the current is proportional to the charge produced by the cell and also to the avalanche triggering probability. Since the produced charge is proportional to V, and we assume that in this region the triggering probability is also a linear function of V, the quadratic relation is obvious for $V > V_{BD}$:

$$I_{SiPM\ cell} \propto q_{cell}(V) P_{trig}(V) \propto (V - V_{BD})(V - V_{BD}). \qquad (1)$$

In equation (1), we assumed only one kind of breakdown voltage. However, the avalanche in a microcell exhibits a hysteresis with two distinct voltages: the "turn-on" and "turn-off" voltages [10]. The turn-on $V_{01}$ voltage is where the microcell initiates avalanching and the turn-off $V_{10}$ voltage is where the avalanche fades away.

This $h$ hysteresis can be taken into account as a voltage shift between the $q_{cell}(V)$ and $P_{trig}(V)$ functions. The cell charge becomes zero below $V_{10}$ and the triggering probability emerges from zero at $V_{01}$, so the current of a single SiPM cell will be a shifted quadratic function of V. For $V > V_{01}$:

$$I_{SiPM\ cell} \propto q_{cell}(V) P_{trig}(V) \propto (V - V_{10})(V - V_{01}), \qquad (2)$$

and the avalanche voltage hysteresis is given by

$$h = V_{01} - V_{10}. \qquad (3)$$

Equation (2) is illustrated graphically in Figure 1.

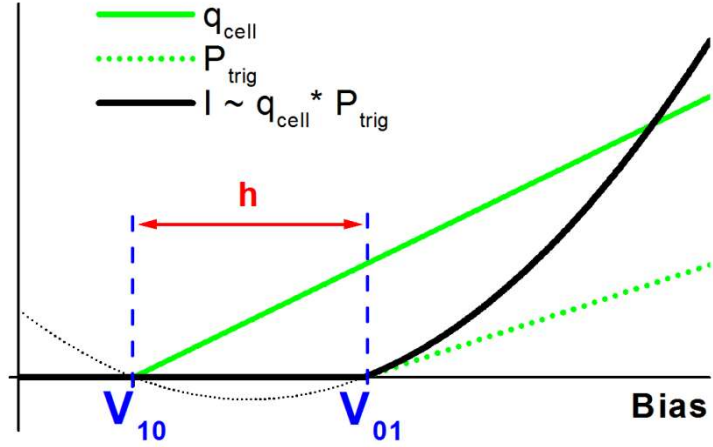

**Figure 1.** The charge produced by a cell above the turn-off $V_{10}$ voltage, and the triggering probability above the turn-on $V_{01}$ voltage are both a linear function of the bias voltage. Thus the cell current becomes a shifted quadratic function of the bias above the turn-on $V_{01}$ voltage. The $h$ hysteresis is the difference between $V_{01}$ and $V_{10}$.

We propose a DC model based on equation (2) and (3). In the DC model we will assume that the microcells in an SiPM have different turn-on voltages forming a distribution $D_{01}(V)$ around the mean turn-on voltage $V_{01}$, but the hysteresis $h$ is common for all microcells. Although the validity of the assumption of a common $h$ cannot be justified a priori, our experimental results shown later in Section 4 are in good agreement with the model predictions. Since the SiPM current is made up of the current of all individual cells, the analytical model of the SiPM current can be expressed by the following integral:

$$I_{SiPM}(V) \propto \int_{-\infty}^{+\infty} D_{01}(v) \cdot (V - v) \cdot (V - v + h) \cdot H(V - v) dv, \qquad (4)$$

where $H(V)$ is the Heaviside step function.

Note that this integral is a convolution of the $D_{01}(V)$ distribution and the single cell current:

$$I_{SiPM}(V) \propto D_{01}(V) * \big(V \cdot (V + h) \cdot H(V)\big). \qquad (5)$$

In order to extract the turn-on voltage distribution $D_{01}(V)$ and the hysteresis $h$ from the convolutional expression of equation (5), we use the 3rd derivative of the SiPM current. Exploiting the relation for the derivative of the convolution, $(f * g)''' = f * g'''$, the 3rd derivative of the SiPM current becomes

$$(I_{SiPM})''' \propto D_{01}(V) * \big(V \cdot (V + h) \cdot H(V)\big)''', \qquad (6)$$

which results in the simple expression

$$(I_{SiPM})''' \propto D_{01}(V) + \frac{h}{2}D_{01}'(V). \qquad (7)$$

Table 2 demonstrates step by step, how we get from equation (6) to equation (7). The first four rows of Table 2 show the steps as the 3$^{rd}$ derivative of a single cell current develops. The turn-on voltage of this particular cell was arbitrarily chosen as $V_{01}$, so the 3$^{rd}$ derivative of its I-V curve will be proportional to a Dirac delta function at exactly $V_{01}$ plus h/2 times the derivative of the same delta function. In the last row of the table a Gaussian turn-on voltage distribution $D_{01}(V)$ was assumed. The sum of this distribution and h/2 times its derivative results in a bipolar shaped 3$^{rd}$ derivative curve, a characteristic shape that we observed for real measured data as shown in the next sections.

**Table 2**: Step by step explanation of the DC model. *H(V)* is the Heaviside step function and $\delta(V)$ is its derivative, the Dirac delta function.

| | |
|---|---|
| **Single cell current:** $$I_{cell} \propto (V - V_{10}) \cdot (V - V_{01}) \cdot H(V - V_{01})$$ | 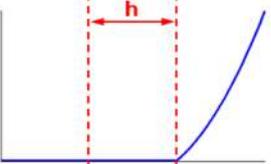 |
| **1$^{st}$ derivative of cell current:** $$I_{cell}' \propto \left[(V - V_{01}) + \frac{h}{2}\right] \cdot H(V - V_{01})$$ | 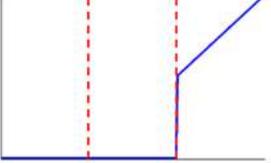 |
| **2$^{nd}$ derivative of cell current:** $$I_{cell}'' \propto H(V - V_{01}) + \frac{h}{2}\delta(V - V_{01})$$ | 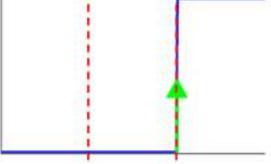 |
| **3$^{rd}$ derivative of cell current:** $$I_{cell}''' \propto \delta(V - V_{01}) + \frac{h}{2}\delta'(V - V_{01})$$ | 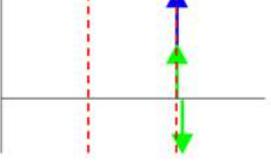 |
| **3$^{rd}$ derivative of SiPM current:** $$I_{SiPM}''' \propto D_{01}(V) + \frac{h}{2}D_{01}'(V)$$ | 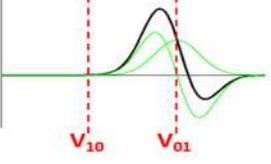 |

# 3. Materials and methods

The measurements were performed on 16 samples (SiPM0 – SiPM15) of the type MPPC S12572-015P from Hamamatsu. This type has a surface of 3x3 mm$^2$, a P-on-N layer order and 40000 microcells.

The sensors under test and the light source were placed in a light-tight box.

For measurements under illuminated conditions, we used a FNL-U501B07WCSL-type, blue (490 nm) LED driven by a current source. The light of the LED was directed to the sensor through a diffuser to distribute the light uniformly.

Current-voltage curves were taken with a 2634B–type Keithley SourceMeter. The default settling times were used in automatic mode. The integration time was set to 10 PLC (Power Line Cycle), that is, 200 ms for 50 Hz power line frequency. A voltage step value of 0.05 V was chosen for the I-V curves. The measurements were performed at 21 °C, with LED illumination. The illumination level was chosen so as the measured current at the recommended operation bias of the SiPM was only 1000 times higher than the dark current. The derivative of the curves was taken 3 times. After the derivations a 11-sample wide, quartic/quintic Savitzky-Golay filter was applied 3 times to obtain a smooth enough curve for the evaluation [11]. The Savitzky-Golay filter was chosen since this filter leaves the mean of the distribution intact and has a small distortion effect on the width and amplitude of the distribution.

The turn-on voltage distribution $D_{01}(V)$ and $h$ hysteresis are not revealed directly by the 3$^{rd}$ derivative of a measured I-V curve as seen in equation (7), which is repeated below for clarity:

$$(I_{SiPM})''' \propto D_{01}(V) + \frac{h}{2} D_{01}{'}(V). \tag{8}$$

However, if we assume a Gaussian distribution for $D_{01}(V)$ then extracting the parameters of $D_{01}(V)$ and the hysteresis $h$ becomes straightforward. Supposing a Gaussian distribution of the turn-on voltages with a mean of $V_{01}$

$$D_{01}(V) = A \cdot exp\left[-\frac{(V - V_{01})^2}{2\sigma^2}\right], \tag{9}$$

the fitting function for the 3$^{rd}$ derivative curves is obtained by plugging equation (9) into equation (7):

$$y = A \cdot \left[2 - \frac{h}{\sigma^2}(V - V_{01})\right] \cdot exp\left[-\frac{(V - V_{01})^2}{2\sigma^2}\right]. \tag{10}$$

The resulting parameters of the fitted functions are the standard deviation ($\sigma$), mean ($V_{01}$), amplitude ($A$) and hysteresis ($h$). Then $V_{10}$ is obtained by subtracting $h$ from $V_{01}$.

Figure 2 shows the I-V curve of SiPM1, its 3rd derivative and the fit by the model from 63.5 V to 66 V. Looking at the 3rd derivative curve, one can observe a bipolar shape as predicted by our model.

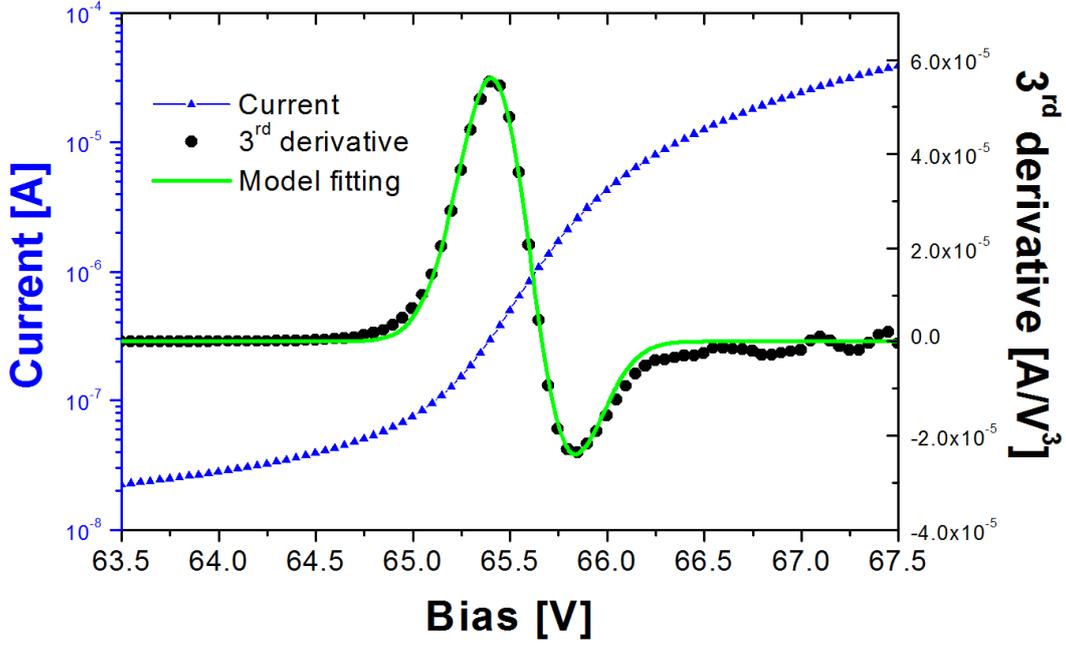

**Figure 2.** The I-V curve of SiPM1, its 3rd derivative and the fit by the model from 63.5 V to 66 V.

## 4. Experimental results

In the first step the current-voltage curves of the 16 SiPM samples were evaluated with the proposed 3rd derivative method. Then the breakdown voltages determined independently from single photon spectrum measurements ($V_{BD}^{SPS}$) were used as a reference to qualify the method. The extracted turn-on $V_{01}$ and turn-off $V_{10}$ voltages of the SiPMs and their relation to $V_{BD}^{SPS}$ values are shown in Table 3. Note that the $V_{10}$ values are very close to their corresponding $V_{BD}^{SPS}$ values and about 1 V below the $V_{01}$ values. This difference between the turn-on and turn-off voltages are the same as the value recently measured [9] for a KETEK SiPM with the same microcell size of 15 μm as in our devices. The Gaussian model of the 3rd derivative method also yields the standard deviation of the breakdown voltages from the fitting function of equation (10). This spread was derived for each SiPM and the value was 0.21 V for all 16 devices.

In Table 3 the factory-given breakdown voltages are also listed for a comparison. Hamamatsu provided the factory data at 25 °C, so the $V_{BD}^{factory}$ values in Table 3 were corrected for 21 °C by a 60 mV/°C temperature coefficient. According to references [4] and [5], to determine the breakdown voltage Hamamatsu may use the "relative derivative" DC method (described in Section 1.), but they combine it with the single photon spectrum (SPS) method as follows. For a certain batch of sensors the difference between the breakdown voltages derived from the relative derivative and SPS method is a constant. This constant is obtained from relative derivative and SPS measurements on some arbitrarily chosen sensors of the batch. Once this constant is determined, it is enough to measure the breakdown voltages by only the relative derivative method for the rest of the batch.

Table 3: The results of the 3$^{rd}$ derivative method calculation for 16 SiPM samples of type MPPC S12572-015P from Hamamatsu. The extracted turn-on $V_{01}$ and turn-off $V_{10}$ voltages of the SiPM are compared to the SPS data $V_{BD}^{SPS}$. The turn-off voltages are in good agreement with the $V_{BD}^{SPS}$ values. The average difference between $V_{01}$ and $V_{10}$ is about 1 V.

| SiPM # | $V_{BD}^{factory}$ [V] | $V_{BD}^{SPS}$ [V] | $V_{10}^{3rd}$ [V] | $V_{01}^{3rd}$ [V] | $V_{10}^{3rd} - V_{BD}^{SPS}$ [V] | $V_{01}^{3rd} - V_{BD}^{SPS}$ [V] |
|---|---|---|---|---|---|---|
| 1 | 65.57 | 65.57 | 65.54 | 66.57 | -0.03 | 1.000 |
| 2 | 64.56 | 64.57 | 64.52 | 65.58 | -0.05 | 1.010 |
| 3 | 65.48 | 65.48 | 65.45 | 66.51 | -0.03 | 1.030 |
| 4 | 64.52 | 64.54 | 64.47 | 65.55 | -0.07 | 1.010 |
| 5 | 65.58 | 65.54 | 65.51 | 66.57 | -0.03 | 1.030 |
| 6 | 65.32 | 65.37 | 65.35 | 66.39 | -0.02 | 1.020 |
| 7 | 64.4 | 64.36 | 64.32 | 65.39 | -0.04 | 1.030 |
| 8 | 65.32 | 65.32 | 65.26 | 66.35 | -0.06 | 1.030 |
| 9 | 65.43 | 65.45 | 65.39 | 66.46 | -0.06 | 1.010 |
| 10 | 64.33 | 64.41 | 64.37 | 65.44 | -0.04 | 1.030 |
| 11 | 64.33 | 64.43 | 64.39 | 65.44 | -0.04 | 1.010 |
| 12 | 64.31 | 64.39 | 64.36 | 65.38 | -0.03 | 0.990 |
| 13 | 64.31 | 64.37 | 64.33 | 65.37 | -0.04 | 1.000 |
| 14 | 64.33 | 64.40 | 64.36 | 65.42 | -0.04 | 1.020 |
| 15 | 64.31 | 64.38 | 64.33 | 65.39 | -0.05 | 1.010 |
| 16 | 64.33 | 64.43 | 64.41 | 65.45 | -0.02 | 1.020 |
| Mean [V]: | | | | | -0.041 | 1.016 |
| Sigma [V]: | | | | | 0.014 | 0.013 |

As a comparison, the same 16 current-voltage curves were also evaluated with all 5 DC methods previously mentioned in Section 1. The mean differences of $V_{BD}^{SPS}$ and the resulting breakdown voltages from the 5 DC methods together with those from the 3$^{rd}$ derivative method are summarized in Table 4. Methods 2, 3, 5 measure a value in statistical accordance with $V_{01}$, while methods 1 and 4 measure a value between $V_{01}$ and $V_{10}$. While the precision of the breakdown voltage values are comparable, it is only the 3rd derivative method that yields the relevant $V_{10}$ with an acceptable accuracy.

**Table 4**: Accuracy (mean) and precision (sigma) of the extracted breakdown voltages by six DC methods. The mean and sigma values were calculated from I-V measurements for 16 SiPM samples of type MPPC S12572-015P from Hamamatsu.

| Method | Mean of ($V_{BD}^{DC}$ - $V_{BD}^{SPS}$) [V] | Sigma [V] |
|---|---|---|
| 1: Tangent | 0.380 | 0.021 |
| 2: Relelative deriv. | 0.938 | 0.021 |
| 3: Inv. rel. deriv. | 0.874 | 0.012 |
| 4: 2nd deriv. | 0.638 | 0.021 |
| 5: Parabolic | 1.016 | 0.045 |
| 6: 3rd deriv. ($V_{01}$) | 1.016 | 0.013 |
| 6: 3rd deriv. ($V_{10}$) | -0.041 | 0.014 |

# 5. Conclusion

We presented a model of the SiPM's current-voltage curve concentrating on the part around its corner point.

- The model takes into account the avalanche turn-on and turn-off voltages and the turn-on voltage distribution over the microcell population.
- It also provides an explanation for the difference between breakdown voltages from single photon spectrum and DC measurements.

Based on the model we developed a new method to determine the SiPM breakdown voltage from the I-V curve.

- The 3rd derivative of the SiPM current yields the distinct turn-on and turn-off voltages.
- The turn-off voltage derived from an illuminated I-V curve is identical to the breakdown voltage determined by single photon spectrum measurements.

We showed that traditional DC methods measure a breakdown voltage either near the turn-on voltage or between the turn-on and turn-off voltages while the new 3rd derivative method gives the relevant turn-off voltage accurately.

# Acknowledgements

The authors wish to thank Tamas Majoros, Faculty of Informatics, University of Debrecen for the single photon spectrum measurements.

References

[1] Roberto Pagano, "OPERATIVE PARAMETERS OF Si PHOTOMULTIPLIERS,", PhD thesis, University of Catania, 2010.

[2] R. Pagano, S. Lombardo, F. Palumbo, D. Sanfilippo, G. Valvo, G. Fallica, and S. Libertino, "Radiation hardness of silicon photomultipliers under 60Co γ-ray irradiation," *Nucl. Instruments Methods Phys. Res. Sect. A Accel. Spectrometers, Detect. Assoc. Equip.*, vol. 767, no. DECEMBER, pp. 347–352, 2014.

[3] E. Garutti, M. Ramilli, C. Xu, and L. Hellweg, "Characterization and X-Ray damage of Silicon Photomultipliers," in *PoS(TIPP2014)070*, 2014, pp. 1–6.

[4] G. Bonanno, D. Marano, M. Belluso, S. Billotta, A. Grillo, S. Garozzo, G. Romeo, and M. C. Timpanaro, "Characterization measurements methodology and instrumental set-up optimization for new sipm detectors - Part II: Optical tests," *IEEE Sens. J.*, vol. 14, no. 10, pp. 3567–3578, 2014.

[5] Hamamatsu Photonics K.K., "MPPC , MPPC modules," pp. 1–39, 2014.

[6] M. Simonetta, M. Biasotti, G. Boca, P. W. Cattaneo, M. De Gerone, F. Gatti, R. Nardò, M. Nishimura, W. Ootani, G. Pizzigoni, M. C. Prata, M. Rossella, N. Shibata, Y. Uchiyama, and K. Yoshida, "Test and characterisation of SiPMs for the MEGII high resolution Timing Counter," *Nucl. Instruments Methods Phys. Res. Sect. A Accel. Spectrometers, Detect. Assoc. Equip.*, vol. 824, pp. 145–147, 2015.

[7] Z. Li, Y. Xu, C. Liu, Y. Li, M. Li, Y. Chen, Y. Wang, B. Lu, W. Cui, J. Huo, T. Chen, D. Han, W. Hu, C. Li, W. Li, X. Liu, J. J. Wang, Y. Yang, Y. Zhang, Y. Zhu, G. Li, J. Zhao, J. J. Wang, N. Pu, and X. Li, "A gain control and stabilization technique for Silicon Photomultipliers in low-light-level applications around room temperature," *Nucl. Instruments Methods Phys. Res. Sect. A Accel. Spectrometers, Detect. Assoc. Equip.*, vol. 695, pp. 222–225, Dec. 2012.

[8] N. Dinu, A. Nagai, and A. Para, "Breakdown voltage and triggering probability of SiPM from IV curves at different temperatures," *Nucl. Instruments Methods Phys. Res. Sect. A Accel. Spectrometers, Detect. Assoc. Equip.*, 2016.

[9] V. Chmill, E. Garutti, R. Klanner, M. Nitschke, and J. Schwandt, "Study of the breakdown voltage of SiPMs," *Nucl. Instruments Methods Phys. Res. Sect. A Accel. Spectrometers, Detect. Assoc. Equip.*, DOI: 10.1016/j.nima.2016.04.047, Apr. 2016.

[10] G. Zhang, D. Han, C. Zhu, and X. Zhai, "Turn-on and turn-off voltages of an avalanche p—n junction," *J. Semicond.*, vol. 33, no. 9, p. 094003, Sep. 2012.

[11] A. Savitzky and M. J. E. Golay, "Smoothing and Differentiation of Data by Simplified Least Squares Procedures.," *Anal. Chem.*, vol. 36, no. 8, pp. 1627–1639, Jul. 1964.